\documentclass[epj,final]{svjour}

\usepackage{times}
\usepackage[dvips]{graphics}
\usepackage[dvips]{epsfig}

\begin{document}

\title{Coherent Dark States of Rubidium 87 in a Buffer Gas using Pulsed Laser
Light}

\author{S. Brattke\thanks{e-mail: brattke@pit.physik.uni-tuebingen.de} \and
U. Kallmann
\and W.-D. Hartmann}   \institute{Universit\"at T\"ubingen, Physikalisches
   Institut, Auf der Morgenstelle 14, 72076 T\"ubingen, Germany}

\abstract{%
The coherent dark resonance between the hyperfine levels $F=1$, $m_F=0$ and
$F=2$, $m_F=0$ of the rubidium ground state has been
observed experimentally with the light of a pulsed mode-locked diode laser
operating at the D1 transition frequency. The resonance
occurs whenever the pulse repetition frequency matches an integer
fraction of the rubidium 87 ground state hyperfine splitting of 6.8 GHz.
Spectra have been taken by varying the pulse repetition frequency. Using
cells with argon as a buffer gas a linewidth as narrow as 149 Hz was obtained.
The rubidium ground state decoherence cross section 
$\sigma_2=1.1\cdot 10^{-18}\,\mathrm{cm}^2$ for collisions with xenon
atoms has been measured for the first time with this method using a pure
isotope rubidium vapor cell and xenon as a buffer gas.}



\maketitle

\section{Introduction}

In the last few years there has been an increasing interest in experiments
on coherent population trapping (CPT) or ``dark resonances'' because of the
variety of interesting effects connected with CPT like
``electromagnetically induced transparency'', ``lasing without inversion''
and precision experiments using the ground state hyperfine structure of
alkali atoms. The topic has been reviewed recently by Arimondo
\cite{arimondo:96}. Recently Brandt et al.\ \cite{brandt:97}
measured a linewidth below 50 Hz by means of a cw-laser CPT experiment 
using cesium vapor and two diode lasers which were phase
locked to each other. CPT experiments can also be performed with
a pulsed laser where the pulse repetition frequency
matches an integer fraction of the ground state splitting. This method was
used first in a time-resolved pump-and-probe experiment by Mlynek et
al.\ \cite{mlynek:81} and later in an experiment
where the pulse repetition frequency of a pulsed diode laser was scanned
over an integer fraction of the rubidium ground state splitting while the
fluorescence light was observed \cite{harde:84,kattau:90}. Using helium
and argon as a buffer gas, linewidths down to 52 Hz were reported. Our work
shows that CPT experiments similar to the one in \cite{harde:84} can
be used to study the interaction of rubidium 87 atoms with heavy noble gas
atoms serving as a buffer gas. We find that the linewidth of the dark
resonance strongly depends on the kind and on the pressure of the buffer gas.
The results are consistent with conventional radio-frequency measurements
\cite{happer:72}. For
the first time using our method we measure the $^{87}\mathrm{Rb}$ ground state
decoherence cross section for collisions with xenon atoms and the pressure
shift of the (0--0) transition caused by xenon as a buffer gas. Further we
could confirm the extreme linewidth reduction with an appropriate choice
of the kind and pressure of the buffer gas, as is shown in 
\cite{brandt:97,harde:84,kattau:90}.

\begin{figure}
\epsfig{figure=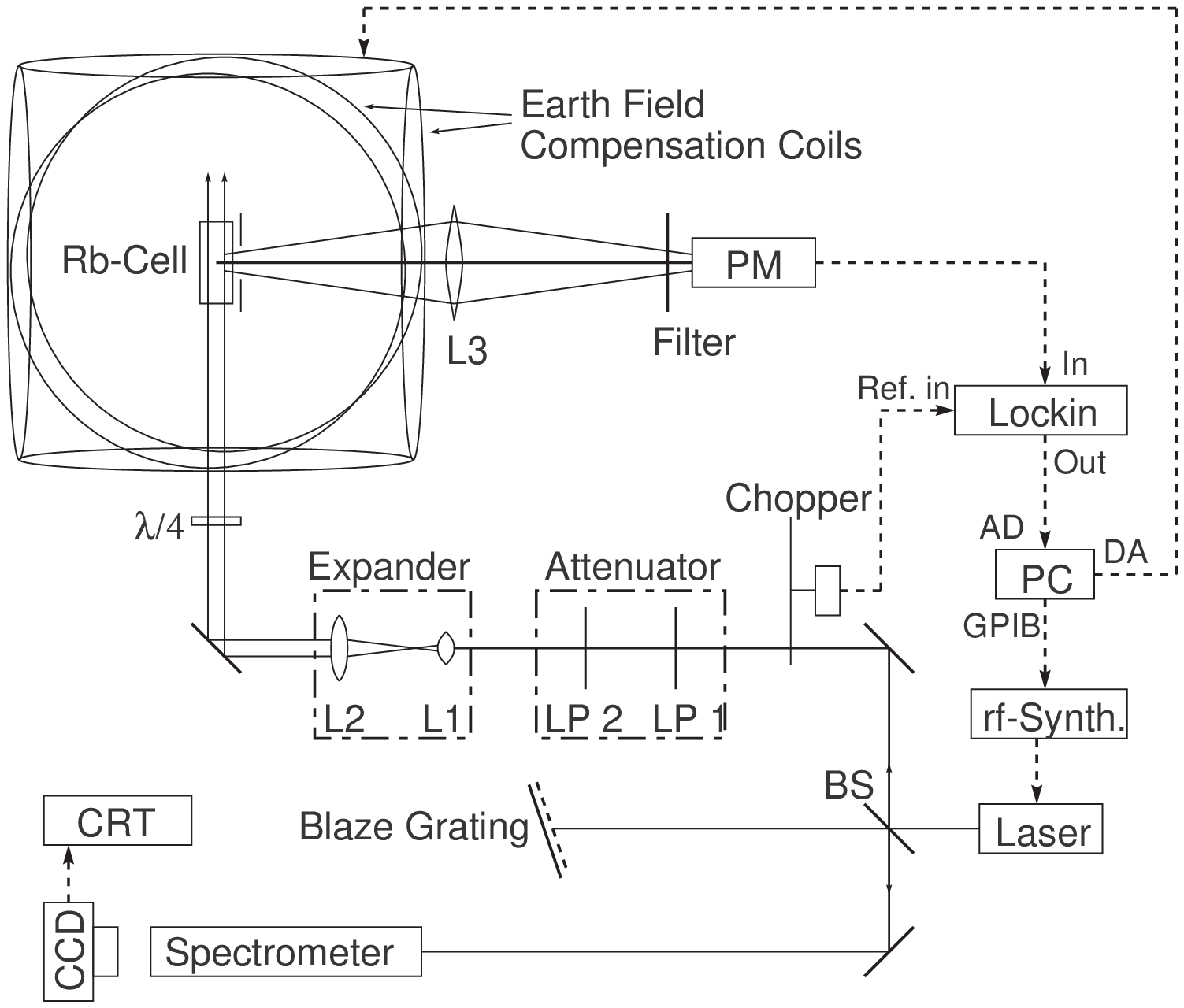,width=8.5cm}
\caption{Experimental setup. L=Lens, LP=Linear polariser, BS=Beam splitter,
PM=photomultiplier.}
\label{fig1}
\end{figure}

\section{Experiment}

\begin{figure}
\epsfig{figure=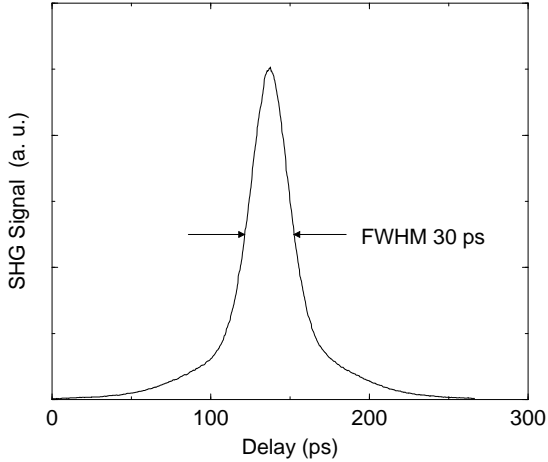,width=8.5cm}
\caption{Autocorrelation of typical pulses}
\label{fig2}
\end{figure}

The experimental setup is shown in figure \ref{fig1}.
We use a SONY SLD 201 laser diode with a ``home made'' anti-reflection
coating on one side in an external cavity with a mode separation of 
525.7 MHz. Pulsed mode-locked operation can be achieved by modulating the
laser current
at this frequency. The light pulses have a duration of 15 ps (FWHM) and 
a spectral width of approximately 100 GHz (FWHM) which is about 85 GHz
bigger than the Fourier limit for such pulses. The pulse width and
shape can be monitored by means of autocorrelation measurements using a SHG
correlator. The autocorrelation of typical pulses as used in the experiment
is shown in fig.\ \ref{fig2}. In order to monitor the laser operation 
on-line we observe the laser spectrum with a grating
spectrometer and a CCD camera. Typically the
information obtained by the spectrum is sufficient to optimize the mode
locking while autocorrelation measurements are only necessary for
reference purposes. The pulse repetition frequency can be scanned over
several tens of kilohertz by varying the modulation frequency of the laser
diode without affecting the laser spectrum and pulse width. The laser is
operating at 795 nm, the D1 transition of rubidium. Transitions out of both
hyperfine ground state levels are induced because of the large spectral width
of the laser.

In order to minimize power broadening and to increase the mean interaction
time of the rubidium atoms with the laser light the chopped beam passes an
attenuator and a beam expander. Then the light is circularly
polarized and the fluorescence light from the rubidium vapor cell
is detected by a photomultiplier with lock-in detection.
The lock-in bandwidth is 1/3 Hz at a chopper
frequency of $22.8\,\mathrm{Hz}$. The pulse repetition frequency is scanned
over an integer fraction of the
ground state level splitting. In our case the mean pulse repetition
frequency of $525.7\,\mathrm{MHz}$ corresponds to 1/13 of the hyperfine
splitting of $6.835\,\mathrm{GHz}$.

The cell is in the center of three mutually perpendicular Helmholtz coil
pairs. Two of these pairs are used to compensate static transversal fields.
The field component parallel to the beam can be adjusted by means of the
third Helmholtz coil. This field causes a Zeeman splitting of the ground
state hyperfine levels. Typically we used fields up to $100\,\mu\mathrm{T}$. 
Fields of this strength are sufficient to Zeeman shift the
magnetic sublevels with $m_{\mathrm{F}}\neq 0$ far enough that only
the (0--0) transition, i.\ e.\ the transition between
the levels $F=1$, $m_{\mathrm{F}}=0$ and $F=2$, $m_{\mathrm{F}}=0$ of
$5^2\mathrm{S}_{1/2}$ is in resonance with the 
the pulse repetition frequency. The 
levels with $m_F=0$ experience only very
small shifts in the applied magnetic field and therefore line broadening
due to inhomogeneities of this field is negligible. 

\begin{figure} 
\begin{center} \epsfig{figure=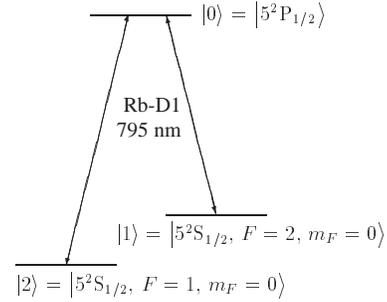,width=5cm} \end{center}
\caption{Rubidium 87 as a $\Lambda$-system}
\label{fig3}
\end{figure}

\section{Results}

We consider the interaction between the atoms and a train of short pulses
resonant to the  rubidium D1 transition.
For our purposes we describe the $^{87}$Rb atom as a three level
$\Lambda$-system (fig.\ \ref{fig3}).
It can be shown theoretically \cite{mlynek:81} that a coherence between the
ground state levels builds up when the pulse repetition frequency matches
a subharmonic of the level splitting frequency. 
When the pulse repetition frequency $\nu_{\mathrm{p}}$ is scanned over
the $m$-th fraction of the hyperfine splitting frequency $\nu_{12}$ the
fluorescence signal is given approximately by the Lorentzian 
\cite{brattke:98}
\begin{equation}
S(m\nu_{\mathrm{p}})\sim
\frac{c_{12}/\Gamma_{12}}{1+\displaystyle
\left(\frac{2\pi(\nu_{12}-m\nu_{\mathrm{p}})}{\Gamma_{12}} \right)^2}
\label{eqn:signal}
\end{equation}
where $\Gamma_{12}$ is the ground state coherence relaxation rate. So
the linewidth only depends on $\Gamma_{12}$ that reads \cite{vanier:89i}
\begin{equation}
\Gamma_{12}=AD_0\frac{p_0}{p}+N_0\bar{v}_r\sigma_2\frac{p}{p_0}
\label{eqn:relax}
\end{equation}
in the presence of a buffer gas at pressure $p$. $A$ is a factor that only
depends on the geometry of the cell, $D_0$ is the diffusion constant of
the rubidium in the buffer gas,  $p_0=1013\,\mathrm{mbar}$ is the reference
pressure, $N_0$ is Loschmidt's constant, $\bar{v}_r$ is the relative
velocity of the rubidium and the buffer gas atoms and $\sigma_2$ is
the decoherence cross section, i.~e.~the cross section for collisions
producing a loss of coherence in the ensemble. The first
term of (\ref{eqn:relax}) describes the relaxation caused by the diffusion of
the rubidium atoms through the buffer gas to the cell walls and the second
term describes decoherence by collisions between rubidium atoms and buffer
gas atoms.

\begin{figure}
\epsfig{figure=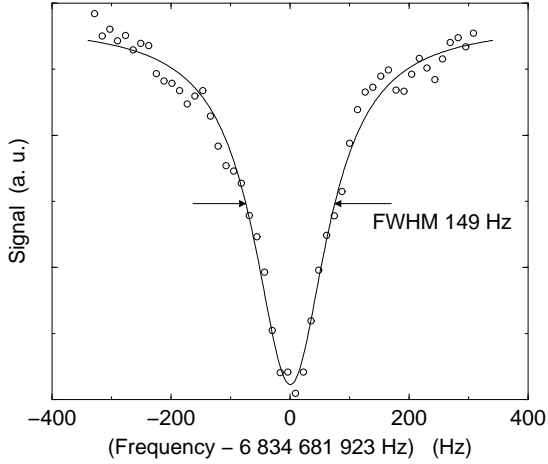,width=8.5cm}
\caption{Typical (0--0) dark resonance in $33\,\mathrm{mbar}$
argon as a buffer gas at $34^\circ$C. The longitudinal magnetic field is
$B=84.1\,\mu\mathrm{T}$. The solid line is a fitted Lorentzian.} 
\label{fig4}
\end{figure}

\begin{figure}
\epsfig{figure=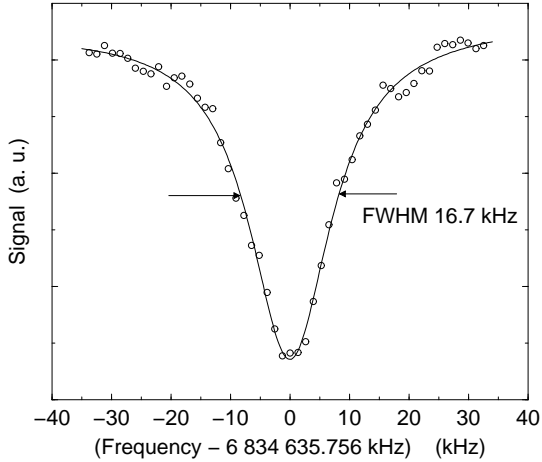,width=8.5cm}
\caption{Typical (0--0) dark resonance in $53\,\mathrm{mbar}$
xenon as a buffer gas at $21^\circ$C. The longitudinal magnetic field is
$B=50\,\mu\mathrm{T}$. The solid line is a fitted Lorentzian.}
\label{fig5}
\end{figure}

We use cylindrical cells with a length of $5\,\mathrm{cm}$ and a diameter of
$2\,\mathrm{cm}$. Using $33\,\mathrm{mbar}$ of argon as a buffer gas in a cell
containing rubidium in the natural isotopic mixture we could measure the
dark resonance with linewidths (FWHM) as narrow as
$149\,\mathrm{Hz}$. Experimental data and a fitted Lorentzian are shown in
fig.\ \ref{fig4}. The linewidth increases with laser
intensity which is consistent with the results in
\cite{brandt:97,kattau:90}. From (\ref{eqn:signal}) and (\ref{eqn:relax}) it
can be expected that the smallest achievable linewidth using our cells is
about $30\,\mathrm{Hz}$. The larger linewidth we get experimentally is due to
power broadening, the rather poor beam quality of our laser and possibly to
jitter in the output of our rf-synthesizer. Broadening due to inhomogeneities
of the magnetic field can be neglected because we only consider coherences
between levels with $m_{\mathrm{F}}=0$.
Using $53.3\,\mathrm{mbar}$ of xenon as a buffer gas in a pure isotope
rubidium cell allowed us to measure the rubidium ground state decoherence
cross section \cite{vanier:74} for collisions with xenon atoms and the
pressure shift \cite{happer:72} of the (0--0) line caused mainly by 
Van der Waals interaction with the xenon atoms. The experiment yielded a
linewidth of $16.7\,\mathrm{kHz}$ at $21^\circ\mathrm{C}$ (fig.\ \ref{fig5}).
Since the diffusion of rubidium atoms through the xenon buffer gas is very
slow only the second term  in (\ref{eqn:relax}) contributes considerably to
the linewidth. Therefore the result can be considered as a direct measurement
of the decoherence cross section $\sigma_2$. We get
$\sigma_2=1.1\cdot 10^{-18}\,\mathrm{cm}^2$ which
is about one order of magnitude greater than the spin disorientation cross
section $\sigma_1=1.3\cdot 10^{-19}\,\mathrm{cm}^2$ given by Franz
\cite{franz:65}. This is in accordance with experimental results for these
cross sections obtained for $^{85}$Rb in He, Ne and Ar \cite{vanier:74} where
the $\sigma_2$ cross sections turned out to be up to three orders of
magnitudes greater than the respective $\sigma_1$ values for $^{87}$Rb
given in \cite{franz:65}. 
Since we own only one sealed off resonance cell containing enriched 
$^{87}$Rb and Xe buffer gas we cannot measure the pressure shift by
varying the xenon pressure. This shift was obtained by correcting the
resonance frequency of figure \ref{fig5} for the applied magnetic field and
subtracting the hyperfine splitting of the free $^{87}$Rb atom
\cite{vanier:89i}. This results in a pressure shift caused by xenon of
$(-885\pm 128)\,\mathrm{Hz/mbar}$. The uncertainty of this value is mainly
due to the uncertainty of the exact value of the buffer gas pressure at the
time of sealing. The shift compares well with measurements
of the frequency shift of $^{87}$Rb in Ar and Kr \cite{bender:58} and the
frequency shift  of $^{133}$Cs in Ar, Kr and Xe \cite{arditi:61}.
We checked the accuracy of our frequency reference (HP-5340A) by 
calculating the ground state hyperfine splitting of $^{87}$Rb from 
measurements of dark resonances with the buffer gas argon (e.~g.~figure
\ref{fig4}) taking into account the well-known pressure shift of 
$-51\,\mathrm{Hz/Torr}$ \cite{bender:58} for argon and the shift due to the
magnetic field and obtained agreement with the value for the hyperfine
splitting taken from literature \cite{vanier:89i} within $200\,\mathrm{Hz}$.
Thus the frequency measurement contributes only a very small part to the
error of the xenon pressure shift.

Our results show that experiments on dark resonances using pulsed laser
light can be applied to study the interaction of alkali atoms with buffer gas
atoms. The advantage of this kind of experiment compared to conventional
rf measurements and to common CPT experiments using cw-lasers is the
relatively simple experimental setup. In our experiment both broadening and
shift of the (0--0) dark resonance line of rubidium in xenon buffer gas could
me measured. The linewidth yielded the rubidium ground state decoherence
cross section for collisions with xenon atoms.

\end{document}